\documentclass[11pt,a4paper]{article}
\usepackage{jheppub}
\usepackage[T1]{fontenc}
\usepackage{amssymb}
\usepackage{mathtools}
\usepackage{amsmath}
\usepackage{bm}
\usepackage{stackengine}
\usepackage{scalerel}
\usepackage{framed}
\usepackage{mathtools}

\usepackage{color}
\usepackage{bbold}
\usepackage{bm}
\usepackage{latexsym}
\usepackage{braket}
\usepackage{slashed}
\numberwithin{equation}{section}

\makeatletter
\newcommand{\doublewidetilde}[1]{{%
		\mathpalette\double@widetilde{#1}%
}}
\newcommand{\double@widetilde}[2]{%
	\sbox\z@{$\m@th#1\widetilde{#2}$}%
	\ht\z@=.9\ht\z@
	\widetilde{\box\z@}%
}
\makeatother
\newcommand\reallywidetilde[1]{\ThisStyle{%
		\setbox0=\hbox{$\SavedStyle#1$}%
		\stackengine{-.1\LMpt}{$\SavedStyle#1$}{%
			\stretchto{\scaleto{\SavedStyle\mkern.2mu\sim}{.5467\wd0}}{0.5\ht0}%
		}{O}{c}{F}{T}{S}%
}}

\title{\boldmath Celestial conformal blocks of  massless scalars and analytic continuation of the Appell function $F_1$\unboldmath}

\author{Wei Fan}
\affiliation{Department of Physics, School of Science, Jiangsu University of Science and Technology,
	Zhenjiang, 212100, China}

\emailAdd{fanwei@just.edu.cn}

\abstract{
	In celestial conformal field theory (CCFT), the 4d massless scalars are represented by 2d conformal operators with conformal dimensions $h=\bar{h}=(1+i\lambda)/2$. The Mellin transform of 4d massless scalar amplitudes gives the conformal correlators of  CCFT. We study the conformal block decomposition of these celestial correlators in CCFT and obtain the explicit blocks. This conformal block decomposition is highly nontrivial, even for the simplest 4d massless scalar amplitude. We use the analytic continuation of the Appell hypergeometric function $F_1$ and  the method of monodromy projection of conformal blocks, to achieve this block decomposition. This procedure is consistent with the crossing symmetry, in both the correlator-level and each explicit block-level. We also investigate its behavior in the conformal soft limit and find that the Appell hypergeometric function $F_1$ does not reduce to the Gauss hypergeometric function. This is  different from the block decomposition of celestial gluons we studied before, where the Appell hypergeometric function $F_1$ reduces to  the Gauss hypergeometric function. This difference comes from the shift of conformal dimensions and is the reason why we adopt the new method here for the block decomposition of celestial massless scalars. 
}

\begin{document}
	
	\maketitle
	\flushbottom

	\section{Introduction}
	\label{sec:intro}
	
	Celestial conformal field theory (CCFT)  is a flat holography~\cite{Strominger:2017zoo,Pasterski:2021raf, Pasterski:2021rjz,Raclariu:2021zjz} of Minkowski spacetime.  The bulk QFT in Minkowski spacetime $\mathbb{R}^{1,3}$ can be recast as a boundary CFT in the celestial sphere. The amplitudes are converted from the  momentum basis to the  boost basis (the so-called conformal primary wavefunction)~\cite{Pasterski:2017kqt} and then become the CFT correlation functions. For massless scalars, the celestial correlation function is the Mellin transform of the standard 4d scattering amplitude. Each 4d massless scalar corresponds to a scalar conformal operator in CCFT, belonging to the principal continuous series~\cite{Pasterski:2017kqt,Mack:1974jjo} with conformal dimensions  $h=\bar{h}=(1 + i\lambda)/2, \lambda\in \mathbb{R}$. 
	
	Understanding the spectrum of primary fields is important to reveal the detailed model of CCFT. In standard CFT~\cite{DiFrancesco:1997nk,Duffin:notes}, the conformal block decomposition of  correlation functions is the common tool to bootstrap the spectrum. Recently there has been studies on the conformal block decomposition~\cite{Lam:2017ofc,Nandan:2019jas,Law:2020xcf,Fan:2021isc,Fan:2021pbp,Fan:2022vbz,Fan:2022kpp,Atanasov:2021cje,Jorge-Diaz:2022dmy,Hu:2022syq,De:2022gjn,Garcia-Sepulveda:2022lga,Chang:2022jut} in CCFT. The detailed technique to compute the blocks in CCFT is different from that of the standard CFT. In standard CFT, conformal blocks are computed by gluing  3pt correlators into  4pt correlators that depend on the cross ratio $z\in\mathbb{C}$. The block decomposition is a product of two Gauss hypergeometric functions with arguments of holomorphic $z$ and antiholomorphic $\bar{z}$ respectively. Since CCFT correlators originate from 4d scattering amplitudes, all 3pt correlators vanish due to 4d momentum constraints. Celestial 4pt correlators contain a delta function  $\delta(z-\bar{z})$ enforcing the cross ratio to be a real variable $z\in\mathbb{R}$. 
	
	We apply the shadow transform~\cite{Ferrara:1972xe,Ferrara:1972ay,Ferrara:1972uq,Ferrara:1972kab} to celestial 4pt correlators to replace one of the four conformal operators by its shadow operator. This relaxes the kinematic constraints on celestial 4pt correlators and the cross ratio becomes a complex variable as in standard CFT. Then the block decomposition of the shadowed celestial correlators proceeds as of standard CFT correlators. Even in standard CFT, the shadow transform plays an important role in the block decomposition~\cite{Dolan:2011dv,Osborn:2012vt,SimmonsDuffin:2012uy}. If an operator $\mathcal{O}$ is in a CFT, its shadow $\tilde{\mathcal{O}}$ also appears in that CFT.  So the shadowed correlator belongs to the putative CCFT, as the unshadowed correlator does. Thus these blocks of the shadowed correlators belong to the spectrum of the CCFT of massless scalars. 
	
	Recently, we have studied the conformal block decomposition of celestial gluon amplitudes~\cite{Fan:2021isc,Fan:2021pbp,Fan:2022vbz,Fan:2022kpp} using this method. A connection between the celestial Yang-Mills theory with the 2d Liouville theory~\cite{Stieberger:2022zyk,Taylor:2023bzj,Stieberger:2023fju} is observed from these results. 
	
	Here we study the conformal block decomposition of celestial massless scalars of the 4d massless $\phi^4$ theory. We study the simplest 4pt amplitude---the contact term. However, its conformal blocks turn out to be  \emph{highly nontrivial}. The correlator is given by the Appell hypergeometric function $F_1$, even in the conformal soft limit. This is different from the case of gluon correlators we studied before~\cite{Fan:2021isc,Fan:2021pbp,Fan:2022vbz,Fan:2022kpp} . The 4pt gluon correlator reduces from the Appell function $F_1$ to the Gauss hypergeometric function in the conformal soft limit. Then we can embed it into a Coulomb-gas model~\cite{dotsenko:notes,Dotsenko:1984ad,Dotsenko:1984nm} and obtain the standard conformal block decomposition. Away from the conformal soft limit, the conformal blocks of gluon correlators are computed using the Banerjee-Ghosh differential equation~\cite{Banerjee:2020vnt,Hu:2021lrx}. But we do not have an analog differential equation for the massless scalars. 
	
	Here we resolve this problem using the analytic continuation of the Appell hypergeometric function $F_1$~\cite{Olsson1964,Bezrodnykh2017}. The analytic continuation of $F_1$  introduces the new hypergeometric function $G_2$~\cite{erdelyi1950} and also generates functions whose arguments mix together the holomorphic cross ratio $z$ and the antiholomorphic cross ratio $\bar{z}$. At first glance, this mixture of arguments of $z$ and $\bar{z}$ looks like an obstacle to the conformal block decomposition. To solve this issue, we adopt the monodromy projection method~\cite{SimmonsDuffin:2012uy}. This method clarifies  conformal blocks by their monodromy and discards terms of wrong monodromy from the integration. The functions whose arguments mix $z$ and $\bar{z}$ indeed have the wrong monodromy and are thrown away by the monodromy projection. The remaining Appell functions can be factorized into standard conformal blocks.

	The paper is organized as follows. In Section 2, we compute the 4pt celestial scalar correlator of the 4d massless $\phi^4$ theory and compute its shadowed correlator. In Section 3, we compute the conformal block of the shadowed correlator and check the crossing symmetry of conformal blocks. In Section 4, we introduce the method of the analytic continuation of the Appell function $F_1$ and the monodromy projection. Then we use it to compute the block decomposition for $G_{34}^{21}$. In Section 5, we perform the block decomposition for $G_{32}^{41}$ and $G_{31}^{24}$,  and verify the crossing symmetry for each explicit block. In Section 6, we study the block decomposition in the conformal soft limit. We point out its difference with the block decomposition of celestial gluons. We conclude with a discussion of open questions  in Section 7.

	\section{The 4pt celestial massless scalar amplitude and its shadow}
	
	In CCFT the massless four-momentum $p_i^\mu=\epsilon_i \omega_i q_i^\mu$ is connected with the celestial sphere~\cite{Pasterski:2017kqt} by $q_i =   (1+z_i\bar{z}_i, z_i+\bar{z}_i,  -i(z_i-\bar{z}_i),1-z_i\bar{z}_i)$, where  $\epsilon_i=\pm 1$ represents outgoing/incoming momentum and $\omega_i$ is the energy. 
	In the 4d metric signature $(-+++)$ ~\cite{Pasterski:2017ylz}, the cross ratio is a complex number $z=(z_{12}z_{34})/(z_{13}z_{24})\in\mathbb{C}$ and $\bar{z}$ is its complex conjugate. Each 4d massless scalar corresponds to a scalar conformal operator $\phi_{\Delta_i}(z_i, \bar{z}_i)$ in CCFT with conformal dimensions  $h_i=\bar{h}_i=\Delta_i/2=(1 + i\lambda_i)/2, \lambda_i\in \mathbb{R}$.

	For massless scalars, the corresponding celestial amplitude $\mathcal{A}_n(\{z_i,\bar{z}_i\})$ is a Mellin transform of the 4d n-point amplitude $A_n(p_1,p_2,\ldots,p_n) $
	\begin{align}
		\label{eq:mellin}
		\mathcal{A}_n(\{z_i,\bar{z}_i\})\coloneqq\left\langle \prod_{i=1}^n \phi_{\Delta_i}(z_i, \bar{z}_i) \right\rangle =\displaystyle{\int_0^\infty \prod_{i=1}^n (d\omega_i \omega_i^{i\lambda_i}) A_n(p_1,p_2,\ldots,p_n) \delta^{4}(\sum_{i=1}^n \epsilon_i \omega_i q_i)},
	\end{align}
	where the celestial amplitude is written as a CCFT correlator of conformal operators $\phi_{\Delta_i}(z_i, \bar{z}_i)$. 
	The momentum-conservating delta function imposes a strong constraint on CCFT correlators. Three-point  correlators are zero because 4d 3pt amplitudes are zero as a consequence of kinematic constraints~\cite{Taylor:2017sph}. The 4pt correlators contain a delta function  $\delta(z-\bar{z})$ enforcing planarity of  scattering events \cite{Pasterski:2017ylz}. We relax kinematic constraints by doing a shadow transform on one of the four operators, which leads to standard CFT correlators with complex-valued cross ratio $z$. 
	
	In this paper, we are interested in the conformal block decomposition of celestial massless scalars. To set up the convention, we start from the standard massless $\phi^4$ theory 
	\begin{align}
		\label{eq:4daction}
		S=\int d^4 x\left(\frac{1}{2}(\partial \phi)^2-\frac{g}{4 !} \phi^4\right), 
	\end{align}
	and consider only the simplest 4pt amplitude, i.e., the contact term  
	\begin{align}
		\label{eq:A4}
		A_4\, \delta^4\left(\sum_{i=1}^{4} p_i\right) =-ig(2 \pi)^4 \delta^4\left(\sum_{i=1}^{4} p_i\right).
	\end{align}
	The conformal block decomposition of  its celestial (shadowed) correlator turns out to be highly nontrivial than naively expected.

	\subsection{4d scattering channels}
	
	We firstly perform the Mellin integral to obtain the  4pt  celestial correlator of massless scalars.  
	The evaluation of the momentum-conservating delta function in the Mellin transform depends on the 4d scattering channels $(ij\rightleftharpoons kl)_{\bm{\mathfrak{4}}}$, where $p_{i,j}$ are outgoing and $p_{k,l}$ are incoming. We use superscripts and subscripts 's, t, u' to represent the $(12\rightleftharpoons 34)_{\bm{\mathfrak{4}}}$, $(13\rightleftharpoons 24)_{\bm{\mathfrak{4}}}$, $(14\rightleftharpoons 23)_{\bm{\mathfrak{4}}}$ channels respectively. Notice that, here channels only distinguish between outgoing/incoming momenta, and has nothing to do with the virtual particle propagations. 
	
	In the 4d $(12\rightleftharpoons 34)_{\bm{\mathfrak{4}}}$ channel, $\epsilon_{1,2}=  1, \epsilon_{3,4}=-1$, the cross ratio is $z>1$ and the delta function evaluates into 
	\begin{align}
		\label{eq:deltaSchannel}
		\delta(\sum_{i=1}^4 \epsilon_i \omega_i q_i) &= \frac{i}{4} \frac{\delta(z-\bar{z})}{\omega_4 |z_{13}|^2 |z_{24}|^2}  \delta(\omega_1 - \omega_4 z |\frac{z_{24}}{z_{12}}|^2) \delta(\omega_2 - \omega_4\frac{z-1}{z} |\frac{z_{34}}{z_{23}}|^2) \delta(\omega_3 -  \omega_4(z-1) |\frac{z_{24}}{z_{23}}|^2).
	\end{align}
	The kinematic constraints lead to $\delta(z-\bar{z})$ which enforce the cross ratio to be on the real axis, rather than on the complex plane as of standard CFT.
	After the Mellin integration, the celestial amplitude evaluates into 
	\begin{align}
		\label{eq:ampSchannel}
		\mathcal{A}_4^s &(z_1,\bar{z}_1,\dots)=\frac{g}{4}(2\pi)^5 \delta(\sum_{i=1}^{4}\lambda_i) \prod_{i<j} z_{ij}^{\frac{h}{3}-h_i-h_j} \bar{z}_{ij}^{\frac{\bar{h}}{3}-\bar{h}_i-\bar{h}_j}\nonumber\\
		&\times \delta(z-\bar{z}) z^{i\lambda_1-i\lambda_2} (z-1)^{i\lambda_2+i\lambda_3}   (z\bar{z})^{\frac{1}{3}-i\frac{\lambda_1}{2}+i\frac{\lambda_2}{2}}  ((z-1)(\bar{z}-1))^{\frac{1}{3}+i\frac{\lambda_1}{2}+i\frac{\lambda_4}{2}} , z>1, 
	\end{align}
	where $h=\bar{h}=\sum_{i=1}^{4} h_i$ and the superscript 's' is used to indicate the 4d scattering channel used in the evaluation of the momentum-conservating delta function. 
	
	It is easy to check that in the 4d $(13\rightleftharpoons 24)_{\bm{\mathfrak{4}}}$ channel,  $\epsilon_{1,3}=  1, \epsilon_{2,4}=-1$, the cross ratio is $0<z<1$ and the celestial amplitude evaluates to 
	\begin{align}
		\label{eq:ampTchannel}
		&\mathcal{A}_4^t (z_1,\bar{z}_1,\dots)=\frac{g}{4}(2\pi)^5\delta(\sum_{i=1}^{4}\lambda_i) \prod_{i<j} z_{ij}^{\frac{h}{3}-h_i-h_j} \bar{z}_{ij}^{\frac{\bar{h}}{3}-\bar{h}_i-\bar{h}_j} \nonumber\\
		&\times\delta(z-\bar{z})z^{i\lambda_1-i\lambda_2} (1-z)^{i\lambda_2+i\lambda_3} (z\bar{z})^{\frac{1}{3}-i\frac{\lambda_1}{2}+i\frac{\lambda_2}{2}}   ((z-1)(\bar{z}-1))^{\frac{1}{3}+i\frac{\lambda_1}{2}+i\frac{\lambda_4}{2}},0<z<1.
	\end{align}
	In the 4d $(14\rightleftharpoons 23)_{\bm{\mathfrak{4}}}$ channel,  $\epsilon_{1,4}=  1, \epsilon_{2,3}=-1$, the cross ratio is $z<0$ and the celestial amplitude is 
	\begin{align}
		\label{eq:ampUchannel}
		&\mathcal{A}_4^u (z_1,\bar{z}_1,\dots)=\frac{g}{4}(2\pi)^5\delta(\sum_{i=1}^{4}\lambda_i)\prod_{i<j} z_{ij}^{\frac{h}{3}-h_i-h_j} \bar{z}_{ij}^{\frac{\bar{h}}{3}-\bar{h}_i-\bar{h}_j}\nonumber\\
		&\times\delta(z-\bar{z}) (-z)^{i\lambda_1-i\lambda_2} (1-z)^{i\lambda_2+i\lambda_3}   (z\bar{z})^{\frac{1}{3}-i\frac{\lambda_1}{2}+i\frac{\lambda_2}{2}}  ((z-1)(\bar{z}-1))^{\frac{1}{3}+i\frac{\lambda_1}{2}+i\frac{\lambda_4}{2}},z<0 .
	\end{align}
	
	The celestial correlator should contain all these solutions. We put them together using the absolute value 
	\begin{align}
		\label{eq:ampSTU}
		&\mathcal{A}_4=\left\langle\phi_{\Delta_1}(z_1, \bar{z}_1) \phi_{\Delta_2}(z_2, \bar{z}_2) \phi_{\Delta_{3}}(z_3, \bar{z}_3) \phi_{\Delta_4}(z_4, \bar{z}_4)\right\rangle =\frac{g}{4}(2\pi)^5\delta(\sum_{i=1}^{4}\lambda_i)\prod_{i<j} z_{ij}^{\frac{h}{3}-h_i-h_j} \bar{z}_{ij}^{\frac{\bar{h}}{3}-\bar{h}_i-\bar{h}_j}\nonumber\\
		&\times \delta(z-\bar{z})  |z|^{i\lambda_1-i\lambda_2} |z-1|^{i\lambda_2+i\lambda_3} (z\bar{z})^{\frac{1}{3}-i\frac{\lambda_1}{2}+i\frac{\lambda_2}{2}}  ((z-1)(\bar{z}-1))^{\frac{1}{3}+i\frac{\lambda_1}{2}+i\frac{\lambda_4}{2}} .
	\end{align}
	When computing with this correlator, we need to replace the absolute value with positive combinations of $z$ and $1$, which will indicate the origin of the 4d scattering channel of the correlator. 
	
	\subsection{The shadowed amplitude}
	
	Now we perform the shadow transform on one of the four operators, to get a complex-valued cross ratio $z$ as of standard CFT correlators. 
	The shadow of a primary operator $\phi_{\Delta}(z, \bar{z})$ with conformal dimension $h,\bar{h}$, hence with the scaling dimension $\Delta=h+\bar{h}$ and spin
	$J=h-\bar{h}$, is defined \cite{Osborn:2012vt} as
	\begin{align}
		\label{eq:shadowDef}
		\tilde{\phi}_{\tilde{\Delta}}(z, \bar{z})\coloneqq\reallywidetilde{\phi_{\Delta}(y, \bar{y})}=k_{h, \bar{h}} \int d^2 y(z-y)^{2 h-2}(\bar{z}-\bar{y})^{2 \bar{h}-2} \phi_{\Delta}(y, \bar{y}),    
	\end{align}
	where the constant $ k_{h, \bar{h}}=(-1)^{2(h-\bar{h})}\Gamma(2-2 h)/(\pi \Gamma(2 \bar{h}-1)) $ is chosen in such a way that, for integer or half-integer spin, $\tilde{\tilde\phi}(z,\bar{z})=\phi(z,\bar{z})$. 
	The shadow field $\tilde{\phi}_{\tilde{\Delta}}(z,\bar{z})$ is  a primary operator with conformal dimension $1-h,1-\bar{h}$, hence with $\tilde{\Delta}=2-\Delta$ and  $ \tilde J=-J$. 
	
	For convenience, we replace the third conformal operator by its shadow operator $\tilde{\phi}_{\tilde{\Delta}_{3}}(z_3', \bar{z}_3') \coloneqq \reallywidetilde{\phi_{\Delta_{3}}(z_3, \bar{z}_3)}$ with conformal dimension $h'_3=\bar{h}'_3=(1-i\lambda_3)/2$
	\begin{align}
		\label{eq:shadowAmp}
		& \left\langle\phi_{\Delta_1}(z_1, \bar{z}_1) \phi_{\Delta_2}(z_2, \bar{z}_2) \tilde{\phi}_{\tilde{\Delta}_{3}}(z_3', \bar{z}_3') \phi_{\Delta_4}(z_4, \bar{z}_4)\right\rangle \nonumber\\
		&= k_{h_3, \bar{h}_3}\int \frac{d^2z_3}{(z_3-z_3')^{1-i\lambda_3}(\bar z_3-\bar z_3')^{1-i\lambda_3}} \left\langle\phi_{\Delta_1}(z_1, \bar{z}_1) \phi_{\Delta_2}(z_2, \bar{z}_2) \phi_{\Delta_{3}}(z_3, \bar{z}_3) \phi_{\Delta_4}(z_4, \bar{z}_4)\right\rangle.
	\end{align}
	After this shadow integral, the new cross ratio $w=(z_{12}z_{34})/(z_{13}'z_{24})$ of the shadowed correlator becomes a complex variable as of standard CFT, with the notation $z_{13}'\coloneqq z_1-z_3'$. We postpone the evaluation of the shadow integral until we practically compute its conformal blocks. 
	
	\section{Conformal blocks of the shadowed amplitude and crossing symmetry}
	
	From now on, we will focus on the shadowed correlator~\eqref{eq:shadowAmp} and study its conformal block decomposition. 
	We follow the convention of~\cite{DiFrancesco:1997nk} in defining the conformal blocks. In the operator approach to CFT, the conformal blocks have analog tree-level Feynman diagrams, which resemble the 4d scattering channels. So the conformal blocks also have analog 2d 'scattering channels'  $(12\rightleftharpoons 34)_{\bm{\mathfrak{2}}}$,  $(14\rightleftharpoons 23)_{\bm{\mathfrak{2}}}$ and  $(13\rightleftharpoons 24)_{\bm{\mathfrak{2}}}$ coming from different conformal transformations. 
	
	Notice that we will \emph{omit the constant factor} $k_{h_3, \bar{h}_3}  (2\pi)^5\delta(\sum_{i=1}^{4}\lambda_i) g/4$ of the conformal blocks in the remaining of this paper.  Here we are only interested in the spectrum of blocks, i.e., the conformal dimension of the blocks. Whenever the exact coefficients of the blocks are needed, the omitted constant factor shall be restored. 
	
	For the 2d $(12\rightleftharpoons 34)_{\bm{\mathfrak{2}}}$ channel, we shall perform a conformal transformation to set 
	$z_1\to \infty, z_2 \to 1, z_4\to 0 $  $z_3'\to w=(z_{12}z_{34})/(z_{13}'z_{24})$ and obtain the function
	\begin{align}
		\label{eq:GsBlockDef}
		G_{34}^{21}(w,\bar{w}) &=  \lim_{z_{1}, \bar{z}_{1}\rightarrow \infty}  z_{1}^{\,2 h_{1}} \bar{z}_{1}^{\,2 \bar{h}_{1}} \left\langle\phi_{\Delta_1}(z_1, \bar{z}_1) \phi_{\Delta_2}(z_2, \bar{z}_2) \tilde{\phi}_{\tilde{\Delta}_{3}}(z_3', \bar{z}_3') \phi_{\Delta_4}(z_4, \bar{z}_4)\right\rangle \nonumber\\
		&= \displaystyle \int dz  |z|^{i\lambda_1+i\lambda_2} |z-1|^{i\lambda_1+i\lambda_4} ((z-w)(z-\bar{w}))^{-1+i\lambda_3},
	\end{align}
	where the delta function $\delta(z-\bar{z})$ constrains the integration variable to be on the real axis $z=\bar{z}$. The conformal block decomposition is to split $G_{34}^{21}(w,\bar{w}) $ into product of two Gauss hypergeometric functions with arguments of holomorphic $w$ and antiholomorphic $\bar{w}$ respectively.  
	
	The conformal block decomposition of  $G_{34}^{21}$ is given by the following sum 
	\begin{align}
		G_{34}^{21}(w,\bar{w}) = \sum_{h,\bar{h}} C_{34}^{21}(h,\bar{h})   K_{34}^{21}[h, \bar{h}] (w,\bar{w}),  
	\end{align}
	where $C_{34}^{21}(h, \bar{h})$ are constants. Each conformal block of a primary field with conformal dimension $(h,\bar h)$ has the form \cite{Osborn:2012vt}
	\begin{align}
		\label{eq:OsbornBlock}
		K_{34}^{21}[h, \bar{h}]=\bar{w}^{\bar{h}-\bar{h}_3-\bar{h}_4}{ }_2 F_1\left(\begin{array}{c}
			\bar{h}-\bar{h}_{12}, \bar{h}+\bar{h}_{34} \\
			2 \bar{h}
		\end{array}; \bar{w}\right) w^{h-h_3-h_4}{ }_2 F_1\left(\begin{array}{c}
			h-h_{12}, h+h_{34} \\
			2 h
		\end{array}; w\right)  
	\end{align}
	where $h_{12}=h_1-h_2$ and $h_{34}=h_3-h_4$. Here to use this definition of $K_{34}^{21}[h, \bar{h}]$,  we need to replace $(h_3, \bar{h}_3)$ by the conformal dimension $(h'_3,\bar{h}'_3)$ of the shadowed operator $\tilde{\phi}_{\tilde{\Delta}_{3}}(z_3', \bar{z}_3') $ 
	\begin{align}
		h_{12}=  \frac{i\lambda_1}{2}-\frac{i\lambda_2}{2},\quad
		h'_{34}\coloneqq h'_3-h_4=-\frac{i\lambda_3}{2}-\frac{i\lambda_4}{2}, \quad h'_3+h_4=1-\frac{i\lambda_3}{2}+\frac{i\lambda_4}{2},
	\end{align}
	where the antiholomorphic ones equal the holomorphic ones.
	
	Besides the 2d $(12\rightleftharpoons 34)_{\bm{\mathfrak{2}}}$ channel, there are other 2d channels. They are obtained by applying  different conformal transformations on the correlator. The conformal symmetry of correlators relates these channels, hence the conformal blocks. This is the crossing symmetry of conformal blocks. Before performing the block decomposition, let's firstly check the crossing symmetry for the blocks of the shadowed correlator. 
	The 2d $(14\rightleftharpoons 23)_{\bm{\mathfrak{2}}}$ channel is obtained by switching $z_2$ and $z_4$ of $G_{34}^{21}$. Now the conformal transformation is $z_1\to \infty, z_2 \to 0, z_4\to 1 $  $z_3'\to 1-w$ and we get 
	\begin{align}
		\label{eq:GuBlockDef}
		G_{32}^{41}(1-w,1-\bar{w}) &=  \lim_{z_{1}, \bar{z}_{1}\rightarrow \infty}  z_{1}^{\,2 h_{1}} \bar{z}_{1}^{\,2 \bar{h}_{1}} \left\langle\phi_{\Delta_1}(z_1, \bar{z}_1) \phi_{\Delta_2}(z_2, \bar{z}_2) \tilde{\phi}_{\tilde{\Delta}_{3}}(z_3', \bar{z}_3') \phi_{\Delta_4}(z_4, \bar{z}_4)\right\rangle \nonumber\\
		&= \displaystyle \int dz  |z|^{i\lambda_1+i\lambda_2} |z-1|^{i\lambda_1+i\lambda_4} ((z-w)(z-\bar{w}))^{-1+i\lambda_3}.
	\end{align}
	Similarly, the 2d $(13\rightleftharpoons 24)_{\bm{\mathfrak{2}}}$ channel is obtained by switching $z_1$ and $z_4$ of $G_{34}^{21}$.  Now the conformal transformation is $z_1\to 0, z_2 \to 1, z_4\to \infty $  $z_3'\to 1/w$ and we get
	\begin{align}
		\label{eq:GtBlockDef}
		G_{31}^{24}(\frac{1}{w},\frac{1}{\bar{w}}) &=  \lim_{z_{4}, \bar{z}_{4}\rightarrow \infty}  z_{4}^{\,2 h_{4}} \bar{z}_{4}^{\,2 \bar{h}_{4}} \left\langle\phi_{\Delta_1}(z_1, \bar{z}_1) \phi_{\Delta_2}(z_2, \bar{z}_2) \tilde{\phi}_{\tilde{\Delta}_{3}}(z_3', \bar{z}_3') \phi_{\Delta_4}(z_4, \bar{z}_4)\right\rangle \nonumber\\
		&= (w\bar{w})^{1-i\lambda_3} \displaystyle \int dz  |z|^{i\lambda_1+i\lambda_2} |z-1|^{i\lambda_1+i\lambda_4} ((z-w)(z-\bar{w}))^{-1+i\lambda_3}.
	\end{align}
	They satisfy the following identities, which is the crossing symmetry of conformal blocks
	\begin{align}
		\label{eq:crossingSym}
		G_{34}^{21}(w,\bar{w}) =G_{32}^{41}(1-w,1-\bar{w}),\quad  G_{34}^{21}(w,\bar{w}) = \frac{1}{w^{2h_3'} \bar{w}^{2\bar{h}_3'}} G_{31}^{24}(\frac{1}{w},\frac{1}{\bar{w}}). 
	\end{align}

	\section{The conformal block decomposition of $G_{34}^{21}$}
	
	Now let's carry out the conformal block decomposition of $G_{34}^{21}(w,\bar{w})$ of the shadowed correlator. The integration variable in~\eqref{eq:GsBlockDef} is the cross ratio $z$ before the shadow transform, constrained to be on the real axis. The integration domain can be divided into three parts $(-\infty, 0)$, $(0,1)$ and $(1,\infty)$, representing contributions to the celestial amplitude  from different 4d scattering channels. So we divide the integral into the following three parts and call them  $I_u, I_t, I_s$ respectively
	\begin{align}
		\label{eq:GsblockIstu}
		G_{34}^{21}(w,\bar{w}) = \displaystyle (\underbrace{\int_{-\infty}^{0}}_{I_u} + \underbrace{\int_{0}^{1}}_{I_t}+\underbrace{\int_{1}^{\infty}}_{I_s} ) dz\, |z|^{i\lambda_1+i\lambda_2} |z-1|^{i\lambda_1+i\lambda_4} ((z-w)(z-\bar{w}))^{-1+i\lambda_3}.    
	\end{align}
	
	These integrals evaluate into Appell hypergeometric functions. The Appell hypergeometric series is defined as
	\begin{align}
		\label{eq:F1Series}
		F_1\left(a , b_1, b_2 , c , x, y\right)=\sum_{m, n=0}^{\infty} \frac{(a)_{m+n}(b_1)_m\left(b_2\right)_n}{(c)_{m+n} m ! n !} x^m y^n,\quad \max (|x|,|y|)<1,   
	\end{align}
	where $(a)_k$ is the Pochhammer symbol 
	\begin{align}
		(a)_k\coloneqq \frac{\Gamma(a+k)}{\Gamma(a)}.    
	\end{align}
	Its analytic continuation is the Appell hypergeometric function $F_1$ of two complex variables $x$ and $y$. Here we use the following integral representation~\cite{NIST:DLMF} 
	\begin{align}
		\label{eq:F1Integral}
		F_1\left(a , b_1, b_2 , c  , x, y\right)=\frac{1}{B(a,c-a)} \int_0^1 \frac{u^{a-1}(1-u)^{c-a-1}}{(1-u x)^{b_1}(1-u y)^{b_2}} \mathrm{d} u,\quad \Re c>0,  \Re(c-a)>0. 
	\end{align}
	with the Euler beta function $B(x,y)=\Gamma(x)\Gamma(y)/\Gamma(x+y)$.
	
	Then the three integrals are expressed as the Appell function $F_1$ with arguments $w$, $1-w$, $1/w$ respectively. For $I_s$ the arguments are $w,\bar{w}$
	\begin{align}
		\label{eq:IsF1}
		I_s&= \int_{1}^{\infty} dz\, z^{i\lambda_1+i\lambda_2} (z-1)^{i\lambda_1+i\lambda_4} ((z-w)(z-\bar{w}))^{-1+i\lambda_3}=B(1-i\lambda_2-i\lambda_3, 1+i\lambda_2+i\lambda_4) \nonumber\\
		&\times F_1(1+i\lambda_2+i\lambda_4, 1-i\lambda_3, 1-i\lambda_3, 2-i\lambda_3+i\lambda_4,  w,\bar{w}).
	\end{align}
	For $I_t$ the arguments are $1/w,1/\bar{w}$
	\begin{align}
		\label{eq:ItF1}
		I_t&= \int_{0}^{1} dz\, z^{i\lambda_1+i\lambda_2} (1-z)^{i\lambda_1+i\lambda_4} ((z-w)(z-\bar{w}))^{-1+i\lambda_3}=B(1-i\lambda_2-i\lambda_3, 1+i\lambda_1+i\lambda_2)\nonumber\\
		&\times  (w\bar{w})^{-1+i\lambda_3}  F_1(1+i\lambda_1 + i\lambda_2, 1-i\lambda_3, 1-i\lambda_3, 2+i\lambda_1 - i\lambda_3, \frac{1}{w},\frac{1}{\bar{w}}),
	\end{align}
	For $I_u$ the arguments are $1-w,1-\bar{w}$
	\begin{align}
		\label{eq:IuF1}
		I_u&= \int_{-\infty}^{0} dz\, (-z)^{i\lambda_1+i\lambda_2} (1-z)^{i\lambda_1+i\lambda_4} ((z-w)(z-\bar{w}))^{-1+i\lambda_3}=B(1+i\lambda_2 +i\lambda_4, 1+i\lambda_1+i\lambda_2)\nonumber\\
		&\times  F_1(1+i\lambda_2 +i\lambda_4, 1-i\lambda_3, 1-i\lambda_3, 2+i\lambda_2 - i\lambda_3, 1-w,1-\bar{w}).
	\end{align}
	
	The block decomposition is to factorize $I_{s,t,u}$ into product of two Gauss hypergeometric functions, in the form of ~\eqref{eq:OsbornBlock}.  We need the Burchnall-Chaundy expansion~\cite{BURCHNALL1,BURCHNALL2} of the Appell hypergeometric function to obtain products of two Gauss hypergeometric functions
	\begin{align}
		\label{eq:F1exp}
		F_1\left(a , b_1, b_2 , c  , x, y\right) =\sum_{n=0}^\infty & {(a)_n(b_1)_n(b_2)_n(c-a)_n \over n!(c+n-1)_n (c)_{2n}}\,x^n y^n\nonumber\\ 
		&\times {}_{2}F_1\left({a+n,b_1+n\atop c+2n};x\right) {}_{2}F_1\left({a+n,b_2+n\atop c+2n};y\right)\ .
	\end{align}
	Using this expansion, the integral $I_s$ can be written in the form of conformal block decomposition~\eqref{eq:OsbornBlock}
	\begin{align}
		\label{eq:IsBlocks}
		I_s&= B(1-i\lambda_2-i\lambda_3, 1+i\lambda_2+i\lambda_4) \sum_{n=0}^\infty {(1+i\lambda_2+i\lambda_4)_n(1-i\lambda_3)_n(1-i\lambda_3)_n(1-i\lambda_2-i\lambda_3)_n \over n!(1+n-i\lambda_3+i\lambda_4)_n (2-i\lambda_3+i\lambda_4)_{2n}}\nonumber\\
		&\times w^{n}  {}_{2}F_1\left({1+n+i\lambda_2+i\lambda_4,1+n-i\lambda_3 \atop 2+2n-i\lambda_3+i\lambda_4};w\right)
		\bar{w}^{n}  {}_{2}F_1\left({1+n+i\lambda_2+i\lambda_4,1+n-i\lambda_3 \atop 2+2n-i\lambda_3+i\lambda_4};\bar{w}\right)
	\end{align}
	The primary fields of the blocks of $I_s$ are $h=\bar{h}=1+n-\frac{i\lambda_3 }{2}+ \frac{i\lambda_4}{2}$. These blocks have trivial monodromy around  $w\to e^{i2\pi} w$, that means the conformal blocks $K_{34}^{21}[h, \bar{h}]$ is a single-valued combination of $w$ and $\bar{w}$. 
	
	Now we need to repeat the Burchnall-Chaundy expansion for $I_{u,t}$  to get their block decomposition. However, the arguments of $I_{u,t}$ are $1-w, 1/w$ respectively. This is a serious problem, because the arguments of Gauss hypergeometric functions should be $w,\bar{w}$ for the blocks of $G_{34}^{21}$. We have met similar problems in the block decomposition of  celestial gluon amplitudes~\cite{Fan:2021pbp}. There the integration is the Gauss hypergeometric function ${}_{2}F_1$ and we use their analytic continuation to change their arguments as ${}_{2}F_1(1-w), {}_{2}F_1(1/w) \to {}_{2}F_1(w)$. The analytic continuation of the Gauss hypergeometric function is a standard method in conformal block decomposition of CFT~\cite{dotsenko:notes,DiFrancesco:1997nk}. 
	
	So here we need to use the analytic continuation of the Appell function $F_1$ to change the arguments of $I_{u,t}$ from $1-w, 1/w$  to $w$. However, the Appell function is a two variable function, which is different from the one variable ${}_{2}F_1$. The analytic continuation of $F_1$ will introduce terms whose arguments mix $w,\bar{w}$ together. Such terms is an obstacle to the holomorphic-antiholomorphic factorization~\eqref{eq:OsbornBlock} of $w$ and $\bar{w}$. Yet, these terms happen to have nontrivial monodromies, i.e., they are not single-valued combinations of $w$ and $\bar{w}$.  So let's firstly review the monodromy properties of conformal blocks, before performing analytic continuation of $F_1$.

	\subsection{Monodromy of conformal blocks}
	
	The monodromy property has been used in the work~\cite{SimmonsDuffin:2012uy} to study  the block decomposition of correlators in standard CFT. Using the shadow formalism, a manifestly conformally-invariant integral is obtained that contains the conformal block $g_{\mathcal{O}}$ and its shadow block $g_{\tilde{\mathcal{O}}}$. The block $g_{\mathcal{O}}$  and its shadow block $g_{\tilde{\mathcal{O}}}$ have different monodromy, so the contribution of the block $g_{\mathcal{O}}$ can be extracted by projecting this conformal integral into the eigenspace with the correct monodromy.  This is called the monodromy projection method~\cite{SimmonsDuffin:2012uy}. By this method, we will keep terms with the correct monodromy  when doing the block decomposition in the conformally-invariant celestial integral. 
	
	In the notation of~\cite{SimmonsDuffin:2012uy} , the monodromy of the scalar block $g_{\mathcal{O}}$ under $x_{12}\to e^{i2\pi} x_{12}$ is as following
	\begin{align}
		\label{eq:DuffinMonodromy}
		&g_{\mathcal{O}} \rightarrow e^{2 \pi i \Delta}  g_{\mathcal{O}},
	\end{align}
	where $\mathcal{O}$ is the primary operator of the block $g_{\mathcal{O}}$. 
	
	In this paper, we use the notation $ K_{34}^{21}[h, \bar{h}]$ of ~\cite{Osborn:2012vt} for conformal blocks. The connection between notations ~\cite{Osborn:2012vt} and ~\cite{SimmonsDuffin:2012uy} is 
	\begin{align}
		K_{34}^{21}[h, \bar{h}](w, \bar{w}) = w^{-h_3-h_4} \bar{w}^{-\bar{h}_3-\bar{h}_4} g_{\mathcal{O}}(w, \bar{w}),
	\end{align}
	where $\Delta=h+\bar{h}$ for  $g_{\mathcal{O}}$. 
	After translating notations,  the monodromy property of the block $K_{34}^{21}$ around $w\to e^{i2\pi} w$  becomes
	\begin{align}
		\label{eq:KMonodromy}
		K_{34}^{21}[h, \bar{h}] &\rightarrow e^{2 \pi i (h-h_3-h_4)}e^{2 \pi i (\bar{h}-\bar{h}_3-\bar{h}_4)}   K_{34}^{21}[h, \bar{h}].
	\end{align}
	If we are only interested in the block $K_{34}^{21}[h, \bar{h}]$, we shall select terms of the integral with the correct monodromy and discard terms with wrong monodromies. 
	
	When applied to the shadowed correlator~\eqref{eq:shadowAmp}, one has to replace $h_3, \bar{h}_3$ by $h'_3, \bar{h}'_3$ in the monodromy projection~\eqref{eq:KMonodromy}. We have seen that these blocks $K_{34}^{21}[h, \bar{h}]$ of $I_s$ have trivial monodromy~\eqref{eq:IsBlocks} around  $w\to e^{i2\pi} w$. So in the analytic continuation of $I_{u,t}$ from $1-w, 1/w$  to $w$, we shall only keep terms that are single-valued combinations of $w$ and $\bar{w}$.  
	
	\subsection{Analytic continuation of the Appell $F_1$}
	
	Now let's use analytic continuation to  change the arguments of $I_{u,t}$ from $1-w, 1/w$  to $w$. 
	The Appell hypergeometric function $F_1$ has the following analytic continuations~\cite{Olsson1964,Bezrodnykh2017}
	\begin{align}
		\label{eq:F1toOne}
		\begin{aligned}
			& F_1\left(a, b_1, b_2, c, x, y\right)=\frac{\Gamma(c) \Gamma\left(c-a-b_1-b_2\right)}{\Gamma(c-a) \Gamma\left(c-b_1-b_2\right)} F_1\left(a, b_1, b_2, 1+a+b_1+b_2-c, 1-x, 1-y\right) \\
			& \quad+\frac{\Gamma(c) \Gamma\left(a+b_2-c\right)}{\Gamma(a) \Gamma\left(b_2\right)}(1-x)^{-b_1}(1-y)^{c-a-b_2} F_1\left(c-a, b_1, c-b_1-b_2, c-a-b_2+1, \frac{1-y}{1-x}, 1-y\right) \\
			& \quad+\frac{\Gamma(c) \Gamma\left(c-a-b_2\right) \Gamma\left(a+b_1+b_2-c\right)}{\Gamma(a) \Gamma\left(b_1\right) \Gamma(c-a)}(1-x)^{c-a-b_1-b_2} \\
			& \quad \times G_2\left(c-b_1-b_2, b_2, a+b_1+b_2-c, c-a-b_2, x-1, \frac{1-y}{x-1}\right),
		\end{aligned}
	\end{align}
	and 
	\begin{align}
		\label{eq:F1toInf}
		\begin{aligned}
			& F_1\left(a, b_1, b_2, c, x, y\right)=\frac{\Gamma(c) \Gamma\left(a-b_1-b_2\right)}{\Gamma(a) \Gamma\left(c-b_1-b_2\right)}(-x)^{-b_1}(-y)^{-b_2} F_1\left(1+b_1+b_2-c, b_1, b_2, 1+b_1+b_2-a, \frac{1}{x}, \frac{1}{y}\right)\\
			& \quad+ \frac{\Gamma(c) \Gamma\left(b_2-a\right)}{\Gamma\left(b_2\right) \Gamma(c-a)}(-y)^{-a} F_1\left(a, b_1, 1+a-c, 1+a-b_2, \frac{x}{y}, \frac{1}{y}\right) \\
			& \quad+\frac{\Gamma(c) \Gamma\left(a-b_2\right) \Gamma\left(b_1+b_2-a\right)}{\Gamma(a) \Gamma\left(b_1\right) \Gamma(c-a)}(-x)^{b_2-a}(-y)^{-b_2} G_2\left(1+a-c, b_2, b_1+b_2-a, a-b_2,-\frac{1}{x}, -\frac{x}{y}\right).
		\end{aligned}
	\end{align}
	The $G_2$ function is defined as an analytic continuation of the following series
	\begin{align}
		\label{eq:G2}
		\begin{aligned}
			& G_2\left(a_1, a_2, b_1, b_2, x, y\right)=\sum_{\substack{m,n=0}}^{\infty}\left(a_1\right)_m\left(a_2\right)_n  \left(b_1\right)_{n-m}\left(b_2\right)_{m-n} \frac{x^m y^n}{m ! n !}, \quad|x|<1, |y|<1.
		\end{aligned}    
	\end{align}
	The various $G_2$ functions and various $F_1$ functions form a complete system of solutions for the Appell hypergeometric differential equations~\cite{erdelyi1950}. This is the reason for the appearance of the $G_2$ function in the analytic continuation of $F_1$. Any solution of the Appell hypergeometric differential equations can be written as a linear combination of three linearly independent Appell hypergeomtric functions~\cite{erdelyi1950}. 
	
	Using~\eqref{eq:F1toInf},  the analytic continuation of $I_t$ gives
	\begin{align}
		\label{eq:ItF1BlockGood}
		&I_t= B(1-i\lambda_2-i\lambda_3,-1+i\lambda_3-i\lambda_4) F_1\left(1+i\lambda_2+i\lambda_4,1-i\lambda_3,1-i\lambda_3,2-i\lambda_3+i\lambda_4,w,\bar{w}\right)\nonumber\\
		& + B(1+i\lambda_1+i\lambda_2, i\lambda_4) (-w)^{-1+i\lambda_3}(-\bar{w})^{-i\lambda_4} F_1\left(1-i\lambda_3-i\lambda_4,1-i\lambda_3,i\lambda_2+i\lambda_3,1-i\lambda_4,\frac{\bar{w}}{w},\bar{w}\right) \nonumber\\
		& +B(1-i\lambda_3+i\lambda_4, -i\lambda_4) (-w)^{-1+i\lambda_3-i\lambda_4} G_2\left(i\lambda_2+i\lambda_3,1-i\lambda_3,1-i\lambda_3+i\lambda_4,-i\lambda_4,-w,-\frac{\bar{w}}{w}\right).
	\end{align}
	The first term has the correct monodromy and it is the same Appell function $F_1$ as of $I_s$~\eqref{eq:IsF1}, with different overall constants. So it has the same conformal block decomposition~\eqref{eq:IsBlocks} as $I_s$, with different overall constants. The second and the third term have the wrong monodromy and are discarded acoording to the monodromy projection method. So the mixing of arguments $w,\bar{w}$ in the second and the third term does not affect the conformal block decomposition. 
	
	Similarly, using~\eqref{eq:F1toOne}, the analytic continuation of $I_u$ gives
	\begin{align}
		\label{eq:IuF1BlockGood}
		&I_u= B(1+i\lambda_2+i\lambda_4,-1+i\lambda_3-i\lambda_4) F_1\left(1+i\lambda_2+i\lambda_4,1-i\lambda_3,1-i\lambda_3,2-i\lambda_3+i\lambda_4,w,\bar{w}\right)\nonumber\\
		& + B(1+i\lambda_1+i\lambda_2, i\lambda_4) (w)^{-1+i\lambda_3}(\bar{w})^{-i\lambda_4} F_1\left(1-i\lambda_3-i\lambda_4,1-i\lambda_3,i\lambda_2+i\lambda_3,1-i\lambda_4,\frac{\bar{w}}{w},\bar{w}\right) \nonumber\\
		& +B(1-i\lambda_3+i\lambda_4, -i\lambda_4) (w)^{-1+i\lambda_3-i\lambda_4} G_2\left(i\lambda_2+i\lambda_3,1-i\lambda_3,1-i\lambda_3+i\lambda_4,-i\lambda_4,-w,-\frac{\bar{w}}{w}\right).
	\end{align}
	Again the first term has the correct monodromy and  the same Appell function $F_1$ as of $I_s$~\eqref{eq:IsF1}, with different overall constants. So it also has the same conformal block decomposition~\eqref{eq:IsBlocks} as $I_s$ with different overall constants. The second and the third term have the wrong monodromy and are the same as the wrong monodromy terms in $I_t$~\eqref{eq:ItF1BlockGood}, except a phase factor $(-1)^{-1+i\lambda_3-i\lambda_4}$. 
	
	Combining the blocks of $I_{s}$~\eqref{eq:IsBlocks}, $I_t$~\eqref{eq:ItF1BlockGood} and $I_u$~\eqref{eq:IuF1BlockGood}, we obtain the following conformal block decomposition of  $G_{34}^{21}(w,\bar{w})$
	\begin{align}
		\label{eq:2dSBlocks}
		&G_{34}^{21}(w,\bar{w}) = \mathcal{N}_{34}^{21}  \sum_{n=0}^\infty {(1+i\lambda_2+i\lambda_4)_n(1-i\lambda_3)_n(1-i\lambda_3)_n(1-i\lambda_2-i\lambda_3)_n \over n!(1+n-i\lambda_3+i\lambda_4)_n (2-i\lambda_3+i\lambda_4)_{2n}}\nonumber\\
		&\times w^{n}  {}_{2}F_1\left({1+n+i\lambda_2+i\lambda_4,1+n-i\lambda_3 \atop 2+2n-i\lambda_3+i\lambda_4};w\right)
		\bar{w}^{n}  {}_{2}F_1\left({1+n+i\lambda_2+i\lambda_4,1+n-i\lambda_3 \atop 2+2n-i\lambda_3+i\lambda_4};\bar{w}\right)
	\end{align}
	with the primary fields of the blocks $h=\bar{h}=1+n-\frac{i\lambda_3 }{2}+ \frac{i\lambda_4}{2}$. 
	Using the Euler's reflection formula~\cite{NIST:DLMF} 
	\begin{align}
		\label{eq:eulerReflection}    
		\Gamma(1-z) \Gamma(z)=\frac{\pi}{\sin \pi z}, \quad z \notin \mathbb{Z},
	\end{align}
	the overall constant  $\mathcal{N}_{34}^{21}$ can be simplified as 
	\begin{align}
		\label{eq:constant2dS}
		\mathcal{N}_{34}^{21} &= B(1-i\lambda_2-i\lambda_3, 1+i\lambda_2+i\lambda_4) +B(1-i\lambda_2-i\lambda_3,-1+i\lambda_3-i\lambda_4) \nonumber\\
		&{\quad }+ B(1+i\lambda_2+i\lambda_4,-1+i\lambda_3-i\lambda_4)\nonumber\\
		&=B(1-i\lambda_2-i\lambda_3, 1+i\lambda_2+i\lambda_4)  \frac{S(-i\lambda_3+i\lambda_4)+S(-i\lambda_2-i\lambda_4)+S(i\lambda_2+i\lambda_3)}{S(-i\lambda_3+i\lambda_4)}
	\end{align}
	where $S(x)\coloneqq \sin{\pi x}$ is the phase factor commonly seen in the conformal block decomposition of standard CFT~\cite{DiFrancesco:1997nk,dotsenko:notes}.

	\section{The conformal block decomposition of $G_{32}^{41}, G_{31}^{24}$ and crossing symmetry}
	
	In the previous section, we perform the conformal block decomposition of $G_{34}^{21}(w,\bar{w})$, which has trivial monodromy around $w\approx 0$. By crossing symmetry of conformal blocks, $G_{34}^{21}(w,\bar{w})$ is related with $G_{32}^{41}(1-w,1-\bar{w})$ and $ G_{31}^{24}(1/w,1/\bar{w})$. We have verified the crossing symmetry in the level of integrals~\eqref{eq:crossingSym}. 
	
	In this section, we will verify the crossing symmetry at the level of each explicit conformal blocks. We will perform the conformal block decomposition of $G_{32}^{41}(1-w,1-\bar{w})$ and $ G_{31}^{24}(1/w,1/\bar{w})$ of the shadowed correlator. They have trivial monodromy around $w\approx 1$ and $w\approx \infty$ respectively, which are used to do the monodromy projection. Then $G_{32}^{41}$ and $G_{31}^{24}$ are written in the form of explicit block decomposition. The explicit blocks of $G_{34}^{21}$ and of $G_{32}^{41}$ are connected by switching $2 \leftrightarrow 4$ and $w \to 1-w, \bar{w}\to 1-\bar{w}$. The explicit blocks of $G_{34}^{21}$  and of $ G_{31}^{24}$ are connected by switching $1 \leftrightarrow 4$ and $w \to 1/w, \bar{w}\to 1/\bar{w}$. This is the crossing symmetry of each block. 
	
	\subsection{Block decomposition of $G_{32}^{41}$}
	
	The conformal block decomposition of  $G_{32}^{41}$ is given by the following sum 
	\begin{align}
		G_{32}^{41}(1-w,1-\bar{w}) = \sum_{h,\bar{h}} C_{32}^{41}(h,\bar{h})  K_{32}^{41}[h, \bar{h}] (1-w,1-\bar{w}),  
	\end{align}
	where $C_{32}^{41}(h, \bar{h})$ are constants. Each conformal block of a primary field with conformal dimension $(h,\bar h)$ has the form
	\begin{align}
		\label{eq:OsbornBlockU}
		K_{32}^{41}[h, \bar{h}]=(1-\bar{w})^{\bar{h}-\bar{h}_3-\bar{h}_2}{ }_2 F_1\left(\begin{array}{c}
			\bar{h}-\bar{h}_{14}, \bar{h}+\bar{h}_{32} \\
			2 \bar{h}
		\end{array}; 1- \bar{w}\right) (1-w)^{h-h_3-h_2}{ }_2 F_1\left(\begin{array}{c}
			h-h_{14}, h+h_{32} \\
			2 h
		\end{array}; 1- w\right)  
	\end{align}
	where $h_{14}=h_1-h_4$ and $h_{32}=h_3-h_2$. For the shadowed correlator~\eqref{eq:shadowAmp}, the parameters are 
	\begin{align}
		h_{14}=  \frac{i\lambda_1}{2}-\frac{i\lambda_4}{2},\quad
		h'_{32}\coloneqq h'_3-h_2=-\frac{i\lambda_3}{2}-\frac{i\lambda_2}{2}, \quad h'_3+h_2=1-\frac{i\lambda_3}{2}+\frac{i\lambda_2}{2},
	\end{align}
	where the antiholomorphic ones equal the holomorphic ones.
	
	From ~\eqref{eq:GuBlockDef} and ~\eqref{eq:GsblockIstu}, we have 
	\begin{align}
		G_{32}^{41}(1-w,1-\bar{w})= I_s + I_u + I_t.     
	\end{align}
	The integral $I_u$ can directly be decomposed into $ K_{32}^{41}[h, \bar{h}]$ using the Burchnall-Chaundy expansion~\eqref{eq:F1exp}
	\begin{align}
		\label{eq:2dUIuBlocks}
		&I_u= B(1+i\lambda_2 +i\lambda_4, 1+i\lambda_1+i\lambda_2) \sum_{n=0}^\infty {(1+i\lambda_2+i\lambda_4)_n(1-i\lambda_3)_n(1-i\lambda_3)_n(1-i\lambda_3-i\lambda_4)_n \over n!(1+n+i\lambda_2-i\lambda_3)_n (2+i\lambda_2-i\lambda_3)_{2n}} (1-w)^{n}\nonumber\\
		&\times   {}_{2}F_1\left({1+n+i\lambda_2+i\lambda_4,1+n-i\lambda_3 \atop 2+2n+i\lambda_2-i\lambda_3};1-w\right)
		(1-\bar{w})^{n}  {}_{2}F_1\left({1+n+i\lambda_2+i\lambda_4,1+n-i\lambda_3 \atop 2+2n+i\lambda_2-i\lambda_3};1-\bar{w}\right)
	\end{align}
	with the primary fields of the blocks $h=\bar{h}=1+n+\frac{i\lambda_2}{2}-\frac{i\lambda_3}{2}$.

	The integral $I_{s,t}$ do not have the correct arguments $1-w$. For $I_s$~\eqref{eq:IsF1}, using the analytic continuation~\eqref{eq:F1toOne} we have
	\begin{align}
		\label{eq:2dUIsBlocks}
		&I_s= B(1+i\lambda_2+i\lambda_4,-1-i\lambda_2+i\lambda_3) F_1\left(1+i\lambda_2+i\lambda_4,1-i\lambda_3,1-i\lambda_3,2+i\lambda_2-i\lambda_3,1-w,1-\bar{w}\right)\nonumber\\
		& + B(1-i\lambda_2-i\lambda_3, i\lambda_2) (1-w)^{-1+i\lambda_3}(1-\bar{w})^{-i\lambda_2} F_1(1-i\lambda_2-i\lambda_3,1-i\lambda_3,i\lambda_3+i\lambda_4,1-i\lambda_2,\frac{1-\bar{w}}{1-w},1-\bar{w}) \nonumber\\
		& +B(1+i\lambda_2-i\lambda_3, -i\lambda_2) (1-w)^{-1-i\lambda_2+i\lambda_3} G_2(i\lambda_3+i\lambda_4,1-i\lambda_3,1+i\lambda_2-i\lambda_3,-i\lambda_2,-(1-w),-\frac{1-\bar{w}}{1-w}).
	\end{align}
	The first term has the correct monodromy and  has the same conformal block decomposition~\eqref{eq:2dUIuBlocks} as $I_u$, with different overall constants. The second and the third term have the wrong monodromy and are discarded acoording to the monodromy projection method.
	
	For $I_t$~\eqref{eq:ItF1}, we need to apply the following identity~\cite{Vavasseur1} before doing the analytic continuation
	\begin{align}
		&(-x)^{-\beta}(-y)^{-\beta^{\prime}} F_1\left(1+\beta+\beta^{\prime}-\gamma, \beta, \beta^{\prime}, 1+\beta+\beta^{\prime}-\alpha , \frac{1}{x}, \frac{1}{y}\right) \nonumber\\
		& = (1-x)^{-\beta}(1-y)^{-\beta^{\prime}} F_1\left(\gamma-\alpha, \beta, \beta^{\prime}, 1+\beta+\beta^{\prime}-\alpha , \frac{1}{1-x}, \frac{1}{1-y}\right).
	\end{align}
	Then the analytic continuation~\eqref{eq:F1toInf} of $I_t$ gives
	\begin{align}
		\label{eq:2dUItBlocks}
		I_t&= B(1+i\lambda_1+i\lambda_2,-1-i\lambda_2+i\lambda_3) F_1\left(1+i\lambda_2+i\lambda_4,1-i\lambda_3,1-i\lambda_3,2+i\lambda_2-i\lambda_3,1-w,1-\bar{w}\right)\nonumber\\
		& + B(1-i\lambda_2-i\lambda_3, i\lambda_2) (-(1-w))^{-1+i\lambda_3}(-(1-\bar{w}))^{-i\lambda_2} \nonumber\\
		&{\quad\quad} \times F_1\left(1-i\lambda_2-i\lambda_3,1-i\lambda_3,i\lambda_3+i\lambda_4,1-i\lambda_2,\frac{1-\bar{w}}{1-w},1-\bar{w}\right) \nonumber\\
		& +B(1+i\lambda_2-i\lambda_3, -i\lambda_2) (-(1-w))^{-1-i\lambda_2+i\lambda_3} \nonumber\\
		&{\quad\quad}\times G_2\left(i\lambda_3+i\lambda_4,1-i\lambda_3,1+i\lambda_2-i\lambda_3,-i\lambda_2,-(1-w),-\frac{1-\bar{w}}{1-w}\right).
	\end{align}
	Again the first term has the correct monodromy and  has the same conformal block decomposition~\eqref{eq:2dUIuBlocks} as $I_u$ with different overall constants. The second and the third term have the wrong monodromy and are the same as the wrong monodromy terms in $I_s$~\eqref{eq:2dUIsBlocks}, except a phase factor $(-1)^{-1-i\lambda_2+i\lambda_3}$.

	Combining the blocks of $I_{s}$~\eqref{eq:2dUIsBlocks}, $I_t$~\eqref{eq:2dUItBlocks} and $I_u$~\eqref{eq:2dUIuBlocks}, we obtain the following conformal block decomposition of  $G_{32}^{41}(1-w,1-\bar{w})$
	\begin{align}
		\label{eq:2dUBlocks}
		&G_{32}^{41}(1-w,1-\bar{w}) = \mathcal{N}_{32}^{41}  \sum_{n=0}^\infty {(1+i\lambda_2+i\lambda_4)_n(1-i\lambda_3)_n(1-i\lambda_3)_n(1-i\lambda_3-i\lambda_4)_n \over n!(1+n+i\lambda_2-i\lambda_3)_n (2+i\lambda_2-i\lambda_3)_{2n}} (1-w)^{n}\nonumber\\
		&\times   {}_{2}F_1\left({1+n+i\lambda_2+i\lambda_4,1+n-i\lambda_3 \atop 2+2n+i\lambda_2-i\lambda_3};1-w\right)
		(1-\bar{w})^{n}  {}_{2}F_1\left({1+n+i\lambda_2+i\lambda_4,1+n-i\lambda_3 \atop 2+2n+i\lambda_2-i\lambda_3};1-\bar{w}\right)
	\end{align}
	with the primary fields of the blocks $h=\bar{h}=1+n+\frac{i\lambda_2}{2}-\frac{i\lambda_3}{2}$.
	Using the Euler's reflection formula~\eqref{eq:eulerReflection}, the overall constant  $\mathcal{N}_{32}^{41}$ can be simplified as 
	\begin{align}
		\label{eq:constant2dU}
		\mathcal{N}_{32}^{41} &= B(1+i\lambda_2 +i\lambda_4, 1+i\lambda_1+i\lambda_2) +  B(1+i\lambda_2+i\lambda_4,-1-i\lambda_2+i\lambda_3)\nonumber\\
		&{\quad }+ B(1+i\lambda_1+i\lambda_2,-1-i\lambda_2+i\lambda_3)\nonumber\\
		&=B(1+i\lambda_2 +i\lambda_4, 1-i\lambda_3-i\lambda_4) \frac{S(i\lambda_2-i\lambda_3)+S(i\lambda_1+i\lambda_3)+S(i\lambda_3+i\lambda_4)}{S(i\lambda_2-i\lambda_3)}.
	\end{align}
	Obviously the blocks $G_{34}^{21}(w,\bar{w})$~\eqref{eq:2dSBlocks} and $G_{32}^{41}(1-w,1-\bar{w})$~\eqref{eq:2dUBlocks} are related by switching  $2 \leftrightarrow 4$ and $w \to 1-w, \bar{w}\to 1-\bar{w}$. So the block decomposition is consistent with the crossing symmetry of conformal blocks. 
	
	\subsection{Block decomposition of $G_{31}^{24}$}
	
	The conformal block decomposition of  $G_{31}^{24}$ is given by the following sum 
	\begin{align}
		G_{31}^{24}(\frac{1}{w},\frac{1}{\bar{w}}) = \sum_{h,\bar{h}} C_{31}^{24}(h,\bar{h})   K_{31}^{24}[h, \bar{h}] (\frac{1}{w},\frac{1}{\bar{w}}),  
	\end{align}
	where $C_{31}^{24}(h, \bar{h})$ are constants.
	Each conformal block of a primary field with conformal dimension $(h,\bar h)$ has the form
	\begin{align}
		\label{eq:OsbornBlockT}
		K_{31}^{24}[h, \bar{h}]=(\frac{1}{\bar{w}})^{\bar{h}-\bar{h}_3-\bar{h}_1}{ }_2 F_1\left(\begin{array}{c}
			\bar{h}-\bar{h}_{42}, \bar{h}+\bar{h}_{31} \\
			2 \bar{h}
		\end{array}; \frac{1}{\bar{w}}\right) ( \frac{1}{w})^{h-h_3-h_1}{ }_2 F_1\left(\begin{array}{c}
			h-h_{42}, h+h_{31} \\
			2 h
		\end{array}; \frac{1}{w}\right)  
	\end{align}
	where $h_{42}=h_4-h_2$ and $h_{31}=h_3-h_1$.For the shadowed correlator~\eqref{eq:shadowAmp}, the parameters are 
	\begin{align}
		h_{42}=  \frac{i\lambda_4}{2}-\frac{i\lambda_2}{2},\quad
		h'_{31}\coloneqq h'_3-h_1=-\frac{i\lambda_3}{2}-\frac{i\lambda_1}{2}, \quad h'_3+h_1=1-\frac{i\lambda_3}{2}+\frac{i\lambda_1}{2},
	\end{align}
	where the antiholomorphic ones equal the holomorphic ones.

	From ~\eqref{eq:GtBlockDef} and ~\eqref{eq:GsblockIstu}, we have 
	\begin{align}
		G_{31}^{24}(\frac{1}{w},\frac{1}{\bar{w}}) =(w\bar{w})^{1-i\lambda_3} ( I_s + I_u + I_t).     
	\end{align}
	The integral $I_t$ can directly be decomposed into $K_{31}^{24}[h, \bar{h}]$ using the Burchnall-Chaundy expansion~\eqref{eq:F1exp}
	\begin{align}
		\label{eq:2dTItBlocks}
		&(w\bar{w})^{1-i\lambda_3} I_t= B(1-i\lambda_2-i\lambda_3, 1+i\lambda_1+i\lambda_2) \sum_{n=0}^\infty {(1+i\lambda_1+i\lambda_2)_n(1-i\lambda_3)_n(1-i\lambda_3)_n(1-i\lambda_2-i\lambda_3)_n \over n!(1+n+i\lambda_1-i\lambda_3)_n (2+i\lambda_1-i\lambda_3)_{2n}} \nonumber\\
		&\times  (\frac{1}{w})^{n} {}_{2}F_1\left({1+n+i\lambda_1+i\lambda_2,1+n-i\lambda_3 \atop 2+2n+i\lambda_1-i\lambda_3};\frac{1}{w}\right)
		(\frac{1}{\bar{w}})^{n}  {}_{2}F_1\left({1+n+i\lambda_1+i\lambda_2,1+n-i\lambda_3 \atop 2+2n+i\lambda_1-i\lambda_3};\frac{1}{\bar{w}}\right)
	\end{align}
	with the primary fields of the blocks $h=\bar{h}=1+n+\frac{i\lambda_1}{2}-\frac{i\lambda_3}{2}$.

	The integral $I_{s,u}$ do not have the correct arguments $1/w$. For $I_s$~\eqref{eq:IsF1}, using the analytic continuation~\eqref{eq:F1toInf} we have
	\begin{align}
		\label{eq:2dTIsBlocks}
		&(w\bar{w})^{1-i\lambda_3}I_s= B(1-i\lambda_2-i\lambda_3,-1-i\lambda_1+i\lambda_3) F_1\left(1+i\lambda_1+i\lambda_2,1-i\lambda_3,1-i\lambda_3,2+i\lambda_1-i\lambda_3,\frac{1}{w},\frac{1}{\bar{w}}\right)\nonumber\\
		& + B(1-i\lambda_1-i\lambda_3, i\lambda_1) (-\frac{1}{w})^{-1+i\lambda_3}(-\frac{1}{\bar{w}})^{-i\lambda_1} F_1(1+i\lambda_2+i\lambda_4,1-i\lambda_3,i\lambda_2+i\lambda_3,1-i\lambda_1,\frac{w}{\bar{w}},\frac{1}{\bar{w}}) \nonumber\\
		& +B(1+i\lambda_1-i\lambda_3, -i\lambda_1) (-\frac{1}{w})^{-1-i\lambda_1+i\lambda_3} G_2(i\lambda_2+i\lambda_3,1-i\lambda_3,1+i\lambda_1-i\lambda_3,-i\lambda_1,-\frac{1}{w},-\frac{w}{\bar{w}}).
	\end{align}
	The first term has the correct monodromy and  has the same conformal block decomposition~\eqref{eq:2dTItBlocks} as $I_t$, with different overall constants. The second and the third term have the wrong monodromy and are discarded acoording to the monodromy projection method.
	
	For $I_u$~\eqref{eq:IuF1}, we need to apply the following identity~\cite{Vavasseur1} before doing the analytic continuation
	\begin{align}
		& F_1\left(\alpha, \beta, \beta^{\prime}, 1+\alpha+\beta+\beta^{\prime}-\gamma , 1-x, 1-y\right) \nonumber\\
		& = x^{-\beta} y^{-\beta^{\prime}} F_1\left(1+\beta+\beta^{\prime}-\gamma, \beta, \beta^{\prime}, 1+\alpha+\beta+\beta^{\prime}-\gamma , \frac{x-1}{x}, \frac{y-1}{y}\right)
	\end{align}
	Then the analytic continuation~\eqref{eq:F1toOne} of $I_u$ gives
	\begin{align}
		\label{eq:2dTIuBlocks}
		&(w\bar{w})^{1-i\lambda_3}I_u= B(1+i\lambda_1+i\lambda_2,-1-i\lambda_1+i\lambda_3) F_1\left(1+i\lambda_1+i\lambda_2,1-i\lambda_3,1-i\lambda_3,2+i\lambda_1-i\lambda_3,\frac{1}{w},\frac{1}{\bar{w}}\right)\nonumber\\
		& + B(1-i\lambda_1-i\lambda_3, i\lambda_1) (\frac{1}{w})^{-1+i\lambda_3}(\frac{1}{\bar{w}})^{-i\lambda_1} F_1(1+i\lambda_2+i\lambda_4,1-i\lambda_3,i\lambda_2+i\lambda_3,1-i\lambda_1,\frac{w}{\bar{w}},\frac{1}{\bar{w}}) \nonumber\\
		& +B(1+i\lambda_1-i\lambda_3, -i\lambda_1) (\frac{1}{w})^{-1-i\lambda_1+i\lambda_3} G_2(i\lambda_2+i\lambda_3,1-i\lambda_3,1+i\lambda_1-i\lambda_3,-i\lambda_1,-\frac{1}{w},-\frac{w}{\bar{w}})
	\end{align}
	Again the first term has the correct monodromy and  has the same conformal block decomposition~\eqref{eq:2dTItBlocks} as $I_t$ with different overall constants. The second and the third term have the wrong monodromy and are the same as the wrong monodromy terms in $I_s$~\eqref{eq:2dTIsBlocks}, except a phase factor $(-1)^{-1-i\lambda_1+i\lambda_3}$.

	Combining the blocks of $I_{s}$~\eqref{eq:2dTIsBlocks}, $I_t$~\eqref{eq:2dTItBlocks} and $I_u$~\eqref{eq:2dTIuBlocks}, we obtain the following conformal block decomposition of  $G_{31}^{24}(\frac{1}{w},\frac{1}{\bar{w}})$
	\begin{align}
		\label{eq:2dTBlock}
		&G_{31}^{24}(\frac{1}{w},\frac{1}{\bar{w}})= \mathcal{N}_{31}^{24}   \sum_{n=0}^\infty {(1+i\lambda_1+i\lambda_2)_n(1-i\lambda_3)_n(1-i\lambda_3)_n(1-i\lambda_2-i\lambda_3)_n \over n!(1+n+i\lambda_1-i\lambda_3)_n (2+i\lambda_1-i\lambda_3)_{2n}} \nonumber\\
		&\times  (\frac{1}{w})^{n} {}_{2}F_1\left({1+n+i\lambda_1+i\lambda_2,1+n-i\lambda_3 \atop 2+2n+i\lambda_1-i\lambda_3};\frac{1}{w}\right)
		(\frac{1}{\bar{w}})^{n}  {}_{2}F_1\left({1+n+i\lambda_1+i\lambda_2,1+n-i\lambda_3 \atop 2+2n+i\lambda_1-i\lambda_3};\frac{1}{\bar{w}}\right)
	\end{align}
	with the primary fields of the blocks $h=\bar{h}=1+n+\frac{i\lambda_1}{2}-\frac{i\lambda_3}{2}$.
	Using the Euler's reflection formula~\eqref{eq:eulerReflection}, the overall constant  $\mathcal{N}_{31}^{24}$ can be simplified as 
	\begin{align}
		\label{eq:constant2dT}
		\mathcal{N}_{31}^{24} &= B(1-i\lambda_2-i\lambda_3, 1+i\lambda_1+i\lambda_2) + B(1-i\lambda_2-i\lambda_3,-1-i\lambda_1+i\lambda_3)\nonumber\\
		&{\quad }+ B(1+i\lambda_1+i\lambda_2,-1-i\lambda_1+i\lambda_3) \nonumber\\
		&=B(1-i\lambda_2 -i\lambda_3, 1+i\lambda_1+i\lambda_2) \frac{S(i\lambda_1-i\lambda_3)+S(i\lambda_2+i\lambda_3)+S(i\lambda_3+i\lambda_4)}{S(i\lambda_1-i\lambda_3)}.
	\end{align}
	Obviously the blocks $G_{34}^{21}(w,\bar{w})$~\eqref{eq:2dSBlocks} and $G_{31}^{24}(\frac{1}{w},\frac{1}{\bar{w}})$~\eqref{eq:2dTBlock} are related by switching $1 \leftrightarrow 4$ and $w \to 1/w, \bar{w}\to 1/\bar{w}$. So the block decomposition is consistent   with the crossing symmetry of each block.

	\section{The conformal soft limit}
	
	For the shadowed correlator $\langle\phi_{\Delta_1} \phi_{\Delta_2} \tilde{\phi}_{\tilde{\Delta}_{3}} \phi_{\Delta_4}\rangle$~\eqref{eq:shadowAmp}, we have achieved the conformal block decomposition and verified its crossing symmetry in both the integral-level and the explicit blocks. These blocks belong to the CCFT of massless scalars. But we still want to know what is the conformal blocks of the unshadowed correlator~\eqref{eq:ampSTU}, which is directly connected with the 4d scattering amplitudes. This is possible in the case of  conformal soft limit $\lambda_3\to 0$. From the definition of the massless scalar conformal primary wave and its shadow transform~\cite{Pasterski:2017kqt},  they are the same operator in the conformal soft limit
	\begin{align}
		\label{eq:softOperator}    
		{\displaystyle\phi_{\Delta_3=1}} \coloneqq \lim_{\lambda_3\to 0}  \phi_{1+i\lambda_3} = \lim_{\lambda_3\to 0}  \reallywidetilde{\phi_{1+i\lambda_3}}=  \lim_{\lambda_3\to 0} \tilde{\phi}_{1-i\lambda_3}. 
	\end{align}
	Then the shadowed correlator equals the original correlator  
	\begin{align}
		\label{eq:softCorrelator}    
		\left\langle\phi_{\Delta_1} \phi_{\Delta_2} \phi_{\Delta_{3}=1} \phi_{\Delta_4}\right\rangle \coloneqq \lim_{\lambda_3\to 0} \left\langle\phi_{\Delta_1} \phi_{\Delta_2} \phi_{\Delta_{3}} \phi_{\Delta_4}\right\rangle = \lim_{\lambda_3\to 0}  \left\langle\phi_{\Delta_1} \phi_{\Delta_2} \tilde{\phi}_{\tilde{\Delta}_{3}} \phi_{\Delta_4}\right\rangle,
	\end{align}
	thus their conformal blocks. Here in the conformal soft limit  $\lambda_3\to 0$,  the integrals $I_{s,t,u}$ of  $G_{34}^{21}(w,\bar{w})$~\eqref{eq:GsblockIstu} become
	\begin{align}
		\label{eq:softIntegral}
		I_s&= B(1-i\lambda_2, 1-i\lambda_1) F_1(1-i\lambda_1, 1, 1, 2+i\lambda_4,  w,\bar{w}) \nonumber\\
		I_t&= B(1-i\lambda_2, 1-i\lambda_4) \frac{1}{w\bar{w}}  F_1(1-i\lambda_4, 1, 1, 2+i\lambda_1, \frac{1}{w},\frac{1}{\bar{w}})\nonumber\\
		I_u&=B(1-i\lambda_1, 1-i\lambda_4)  F_1(1-i\lambda_1, 1, 1, 2+i\lambda_2, 1-w,1-\bar{w}).
	\end{align}
	
	Note that they are still  given by the Appell hypergeometric function $F_1$. This is \emph{completely different} from the case of celestial gluons we studied before~\cite{Fan:2021isc,Fan:2021pbp}. In the case of gluons, the Appell function $F_1$ reduces to the Gauss hypergeometric function ${}_{2}F_{1}$ in the conformal soft limit of the shadowed operator. There the monodromoy property of conformal blocks is solved by embeding  into  a Coulomb-gas model~\cite{dotsenko:notes,Dotsenko:1984ad,Dotsenko:1984nm}. The resulting conformal blocks have the correct crossing symmetry. For celestial massless scalars here, we can not copy this procedure used in celestial gluons, due to the presence of the Appell $F_1$.  This is the reason why here we use the analytic continuation of the Appell $F_1$ and the monodromy projection method, for the conformal block decomposition of celestial massless scalars. 
	
	The reason for this difference between the scalar case and the gluon case lies in their conformal dimensions. For scalars $h_i=\bar{h}_i$, the two parameters $b_1=b_2=b$ of the Appell function are equal $F_1(a, b, b, c; x,y)$. In the conformal soft limit $b\to 1$, so the Appell function $F_1(a, 1,1, c; x,y)$ does not get any essential reduction. For gluons $h_i-\bar{h}_i=\pm 1$, this shift of conformal dimension leads to a difference in the two  parameters $b_1-b_2=2$ of the Appell function. In the conformal soft limit $b_2\to 0$, so the Appell function is reduced to the Gauss hypergeometric function $F_1(a, 2,0, c, x,y)={}_{2}F_{1}(a,2,c;x).$
	
	In this conformal soft limit  $\lambda_3\to 0$, the explicit conformal blocks of  $\left\langle\phi_{\Delta_1} \phi_{\Delta_2} \phi_{\Delta_{3}=1} \phi_{\Delta_4}\right\rangle$ are as following
	\begin{align}
		\label{eq:softBlocks}
		G_{34}^{21}(w,\bar{w}) =&B(1-i\lambda_2, 1-i\lambda_1)  \frac{S(i\lambda_4)+S(i\lambda_1)+S(i\lambda_2)}{S(i\lambda_4)}  \sum_{n=0}^\infty {(1-i\lambda_1)_n(1)_n(1)_n(1-i\lambda_2)_n \over n!(1+n+i\lambda_4)_n (2+i\lambda_4)_{2n}}\nonumber\\
		&\times w^{n}  {}_{2}F_1\left({1+n-i\lambda_1,1+n \atop 2+2n+i\lambda_4};w\right)
		\bar{w}^{n}  {}_{2}F_1\left({1+n-i\lambda_1,1+n \atop 2+2n+i\lambda_4};\bar{w}\right)
	\end{align}
	with the primary fields of the blocks $h=\bar{h}=1+n + \frac{i\lambda_4}{2}$. The correlator $\left\langle\phi_{\Delta_1} \phi_{\Delta_2} \phi_{\Delta_{3}=1} \phi_{\Delta_4}\right\rangle$ directly originate from the corresponding 4d scattering amplitudes. So in the CCFT of massless scalars, if the four scalar conformal operators $\Delta_1, \Delta_2, \Delta_3=1, \Delta_4$ are involved in the  $(12\rightleftharpoons 34)_{\bm{\mathfrak{2}}}$ channel,  the scalar conformal operators of dimension $h=\bar{h}=1+n + \frac{i\lambda_4}{2}$ are also present in this CCFT. Similarly, the other 2d channels lead to conformal primary operators of conformal dimension $h=\bar{h}=1+n + \frac{i\lambda_1}{2}$ and $h=\bar{h}=1+n + \frac{i\lambda_2}{2}$. 
	
	Here all the operators from these blocks are scalars. Let's compare with the operators obtained from conformal blocks of celestial gluons~\cite{Fan:2021pbp}. (A): For gluons, the blocks~\cite{Fan:2021pbp} are computed for four gluon operators with scaling dimension $\Delta_1=1, \Delta_2, \Delta_3, \Delta_4$ and spin $J_1=J_2=-1, J_3=J_4=+1$, with the first operator being conformal soft $\lambda_1=0$. The resulting operators are operators with spin: the $(12\rightleftharpoons 34)_{\bm{\mathfrak{2}}}$ channel gives operators with scaling dimension $\Delta=2+J+i \lambda_2$ and spin  $J \geq 0$, the $(14\rightleftharpoons 23)_{\bm{\mathfrak{2}}}$ channel gives operators with scaling dimension $\Delta=J+i \lambda_4$ and spin  $J \geq 2$, the $(13\rightleftharpoons 24)_{\bm{\mathfrak{2}}}$ channel gives operators with scaling dimension $\Delta=J+i \lambda_3$ and spin  $J \geq 2$.
	(B): For massless scalars, we obtain the blocks with conformal soft limit $\lambda_1=  0$, by  translating the results~\eqref{eq:softBlocks} obtained with conformal soft limit $\lambda_3=  0$. Then  for  four scalar operators with scaling dimension $\Delta_1=1, \Delta_2, \Delta_3, \Delta_4$, the blocks give the following scalar operators: the $(12\rightleftharpoons 34)_{\bm{\mathfrak{2}}}$ channel gives operators with scaling dimension $\Delta=2+2n+i \lambda_2$, the $(14\rightleftharpoons 23)_{\bm{\mathfrak{2}}}$ channel gives operators with scaling dimension $\Delta=2+2n+i \lambda_4$, the $(13\rightleftharpoons 24)_{\bm{\mathfrak{2}}}$ channel gives operators with scaling dimension $\Delta=2+2n+i \lambda_3$. We can see that the scalar operator $\Delta=2+i \lambda_2$ is special, because both the celestial gluons and the massless scalars contain a block with this scaling dimension.

	\section{Conclusion}
	
	In this work, we achieved the conformal block decomposition of celestial massless scalars. We took the  celestial 4pt correlator of the simplest 4d amplitude of $\phi^4$ theory, and obtained its shadowed correlator using the shadow transform. We introduced the analytic continuation of the Appell hypergeometric function $F_1$ and used the monodromy projection~\cite{SimmonsDuffin:2012uy}  to obtain the target conformal blocks. At each stage of our method, the crossing symmetry of conformal blocks was confirmed. The resulting conformal blocks have the form of conformal blocks of standard CFT~\cite{Osborn:2012vt,SimmonsDuffin:2012uy}. We discovered scalar primary operators with scaling dimensions of the form $\Delta=2+2n + i\lambda, n\ge 0$.
	
	We studied the block decomposition in the conformal soft limit. The Appell hypergeometric function $F_1$ does not reduce to the Gauss hypergeometric function and this is  different from the block decomposition of celestial gluons we studied before~\cite{Fan:2021isc,Fan:2021pbp}. This is the reason why we adopt the new method  for the block decomposition of celestial massless scalars. However, we can  ask the question whether it is still possible to set up a Coulomb-gas-like model~\cite{dotsenko:notes,Dotsenko:1984ad,Dotsenko:1984nm} for the conformal block decomposition of celestial massless scalars, as the case of celestial gluons? This is because the analytic continuation of the Appell function $F_1$ resembles the analytic continuation of the Gauss hypergeometric function used in the Coulomb-gas-like model of standard CFT.

	\acknowledgments
	Wei Fan is supported in part by the National Natural Science Foundation of China under Grant No.\ 12105121.




	\bibliographystyle{JHEP}
	\bibliography{scalar_block_CCFT}

\end{document}